\documentclass[graybox]{svmult}

\usepackage{mathptmx}       
\usepackage{helvet}         
\usepackage{courier}        
\usepackage{type1cm}        
\usepackage{makeidx}         
\usepackage{graphicx}        
\usepackage{multicol}        
\usepackage[bottom]{footmisc}

\usepackage{subfig}

\usepackage{amsmath,comment,amssymb,cite,mathtools}
\usepackage{tikz}
 
 \usetikzlibrary{arrows}
 \usetikzlibrary{fit}
 \usetikzlibrary{calc}
  \pgfdeclarelayer{background1}
  \pgfdeclarelayer{background2}
  \pgfdeclarelayer{background3}
  \pgfdeclarelayer{background4}
  \pgfdeclarelayer{foreground}
  \pgfsetlayers{background4,background3,background2,background1,main,foreground}
  \tikzset{errorstyle/.style={thick,red,solid}}
  \tikzset{yrefstyle/.style={thick,black,dashed}}
  \tikzset{safetystyle/.style={thick,blue,dotted}}
  \tikzset{funnelstyle/.style={thick,blue,densely dotted}}
  \tikzset{funnelbackground/.style={black!20,opacity=0.5}}
  \tikzset{funneldstyle/.style={thin,blue,dashed}}
  \tikzset{qstyle/.style={green!50!black,ultra thick}}
  \tikzset{qhelpstyle/.style={green!50!black,very thin}}
  \tikzset{funnelfillstyle/.style={blue!20!white,opacity=0.8}}
  \tikzset{safetyfillstyle/.style={blue,opacity=0.1}}
  \tikzset{funnelthinfillstyle/.style={blue!5!white,opacity=0.8}}
  \tikzset{safetythinfillstyle/.style={blue!50!white,opacity=0.1}}

\newcommand{\R}{\ensuremath{{\mathbb R}}}

\newcommand{\AAA}{{\mathcal A}}

\newcommand{\LL}{{\mathcal L}}

\newcommand{\NN}{{\mathcal N}}

\makeindex             

\begin{document}


\title*{Design of Heterogeneous Multi-agent System for Distributed Computation}
\author{Jin Gyu Lee and Hyungbo Shim}
\institute{Jin Gyu Lee \at University of Cambridge, Department of Engineering, Control Group, Trumpington Street, Cambridge CB2 1PZ, United Kingdom, \email{jgl46@cam.ac.uk}
\and Hyungbo Shim \at Seoul National University, Department of Electrical and Computer Engineering, Gwanak-ro 1, Gwanak-gu, Seoul 08826, Korea, \email{hshim@snu.ac.kr}
}
%
%
\maketitle

\abstract{
A group behavior of a heterogeneous multi-agent system is studied which obeys an ``average of individual vector fields'' under strong couplings among the agents.
Under stability of the averaged dynamics (not asking stability of individual agents), the behavior of heterogeneous multi-agent system can be estimated by the solution to the averaged dynamics. 
A following idea is to ``design'' individual agent's dynamics such that the averaged dynamics performs the desired task.
A few applications are discussed including estimation of the number of agents in a network, distributed least-squares or median solver, distributed optimization, distributed state estimation, and robust synchronization of coupled oscillators. 
Since stability of the averaged dynamics makes the initial conditions forgotten as time goes on, these algorithms are initialization-free and suitable for plug-and-play operation. At last, nonlinear couplings are also considered, which potentially asserts that enforced synchronization gives rise to an emergent behavior of a heterogeneous multi-agent system.
}

\setcounter{page}{1}

\section{Introduction}\label{sec:1}

During the last decade, synchronization and collective behavior of a multi-agent system have been actively studied because of numerous applications in diverse areas, e.g., biology, physics, and engineering.
An initial study was about identical multi-agents \cite{olfati2004consensus,moreau2004stability,Ren05,seo2009consensus}, but the interest soon transferred to the heterogeneous case because uncertainty, disturbance, and noise are prevalent in practice.
In this regard, heterogeneity was mostly considered harmful---something that we have to suppress or compensate.
To achieve synchronization, or at least approximate synchronization (with arbitrary precision if possible), against heterogeneity, various methods such as output regulation \cite{Kim11,de2012internal,isidori2014robust,su2014cooperative,casadei2017multipattern,su2019semi}, backstepping \cite{zhang2016almost}, high-gain feedback \cite{delellis2015convergence,montenbruck2015practical,kim2016robustness,panteley2017synchronization}, adaptive control \cite{lee2018practical}, and optimal control \cite{modares2017optimal}, have been applied.
However, heterogeneity of multi-agent systems is a way to achieve a certain task collaboratively from different agents.
From this viewpoint, heterogeneity is something we should design, or, at least, heterogeneity is an outcome of distributing a complex computation into individual agents.

This chapter is devoted to investigating the design possibility of the heterogeneity.
After presenting a few basic theorems which describe the collective behavior of multi-agent systems, we exhibit several design examples by employing the theorems as a toolkit.
A feature of the toolkit is that the vector field of the collective behavior can be assembled from the individual vector fields of each agent when the coupling strength among the agents is sufficiently large.
This process is explained by the singular perturbation theory.
In fact, the assembled vector field is nothing but an average of the agents' vector fields, and it appears as the quasi-steady-state subsystem (or the slow subsystem) when the inverse of the coupling gain is treated as the singular perturbation parameter.
We call the quasi-steady-state subsystem as {\em blended dynamics} for convenience.
The behavior of the blended dynamics is an emergent one if none of the agents has such a vector field.
For instance, we will see that we can construct a heterogeneous network that individuals can estimate the number of agents in the network without using any global information.
Since individuals cannot access the global information $N$, this collective behavior cannot be obtained by the individuals alone.
On the other hand, appearance of the emergent behavior when we enforce synchronization seems intrinsic.
We will demonstrate this fact when we consider nonlinear coupling laws in the later section.
Finally, the proposed tool leads to the claim that the network of a large number of agents is robust against the variation of individual agents.
We will demonstrate it for the case of coupled oscillators.

There are two notions which have to be considered when a multi-agent system is designed.
It is said that the {\em plug-and-play operation} (or, {\em initialization-free}) is guaranteed for a multi-agent system if it maintains its task without resetting all agents whenever an agent joins or leaves the network.
On the other hand, if a new agent that joins the network can construct its own dynamics without the global information such as graph structure, other agents' dynamics, and so on, it is said that the {\em decentralized design} is achieved.
It will be seen that the plug-and-play operation is guaranteed for the design examples in this chapter.
This is due to the fact that the group behavior of the agents is governed by the blended dynamics, and therefore, as long as the blended dynamics remains stable, individual initial conditions of the agents are forgotten as time goes on.
The property of decentralized design is harder to achieve in general.
However, for the presented examples, this property is guaranteed to some extent; more specifically, it is achieved except the necessity of the coupling gain which is the global information.

\section{Strong diffusive state coupling}\label{sec:state}

We begin with the simplest case of heterogeneous multi-agent systems given by
\begin{align}\label{eq:net_state}
\dot{x}_i = f_i(t, x_i) +  k\sum_{j \in \mathcal{N}_i}\alpha_{ij}(x_j - x_i) \quad \in \mathbb{R}^n, \quad i \in \mathcal{N},
\end{align}
where $\mathcal{N} := \{1, \cdots, N\}$ is the set of agent indices with the number of agents, $N$, and $\mathcal{N}_i$ is a subset of $\mathcal{N}$ whose elements are the indices of the agents that send the information to agent $i$.
The coefficient $\alpha_{ij}$ is the $ij$-th element of the adjacency matrix that represents the interconnection graph.
We assume {\em the graph is undirected and connected} in this chapter.
The vector field $f_i$ is assumed to be piecewise continuous in $t$, continuously differentiable with respect to $x_i$, locally Lipschitz with respect to $x_i$ uniformly in $t$, and $f_i(t, 0)$ is uniformly bounded for $t$.
The summation term in \eqref{eq:net_state} is called diffusive coupling, in particular, {\em diffusive state coupling} because states are exchanged among agents through the term.
Diffusive state coupling term vanishes when state synchronization is achieved (i.e., $x_i(t)=x_j(t)$, $\forall i, j$).
Coupling strength, or coupling gain, is represented by the positive constant $k$.

It is immediately seen from \eqref{eq:net_state} that synchronization of $x_i(t)$'s to a common trajectory $s(t)$ is hopeless in general due to the heterogeneity of $f_i$ unless it holds that $\dot s(t) = f_i(t,s(t))$ for all $i \in \NN$.
Instead, with sufficiently large coupling gain $k$, we can enforce approximate synchronization.
To see this, let us introduce a linear coordinate change 
\begin{align}\label{eq:ct1}
\begin{split}
s &= \frac1N \sum_{i=1}^N x_i \quad \in \R^n \\
\tilde z &= (R^T \otimes I_n) {\rm col}(x_1,\cdots,x_N) \quad \in \R^{(N-1)n}
\end{split}
\end{align}
where $R \in \R^{N \times (N-1)}$ is any matrix that satisfies
$$\begin{bmatrix} \frac1N 1_N^T \\ R^T \end{bmatrix} \LL \begin{bmatrix} 1_N & R \end{bmatrix} = \begin{bmatrix} 0 & 0 \\ 0 & \Lambda \end{bmatrix}$$
with a positive definite matrix $\Lambda \in \R^{(N-1) \times (N-1)}$, where $1_N := [1, \cdots, 1]^T \in \R^N$ and $\mathcal{L}:= \mathcal{D} - \mathcal{A}$ is the Laplacian matrix of a graph, in which $\mathcal{A} = [\alpha_{ij}]$ and $\mathcal{D} = \text{diag}(d_i)$ with $d_i = \sum_{j \in \mathcal{N}_i}\alpha_{ij}$.
By the coordinate change, the multi-agent system is converted into the standard singular perturbation form
\begin{align}\label{eq:spform1}
\begin{split}
\dot s &= \frac1N \sum_{i=1}^N f_i(t,s + (R_i \otimes I_n) \tilde z) \\
\frac1k \dot {\tilde z} &= - (\Lambda \otimes I_n) \tilde z + \frac1k (R^T \otimes I_n) {\rm col}(f_1(t,x_1),\cdots,f_N(t,x_N))
\end{split}
\end{align}
where $R_i$ implies the $i$-th row of $R$.
From this, it is seen that $\tilde z$ quickly becomes arbitrarily small with arbitrarily large $k$, and the quasi-steady-state subsystem of \eqref{eq:spform1} is given by
\begin{equation}\label{eq:bl1}
\dot s = \frac1N \sum_{i=1}^N f_i(t,s),
\end{equation}
which we call {\em blended dynamics}\footnote{More appropriate name could be `averaged dynamics,' which may however confuse the reader with the averaged dynamics in the well-known averaging theory \cite{Sanders} that deals with time average.}.
By noting that $x_i = s + (R_i \otimes I_n) \tilde z$, it is seen that the behavior of the multi-agent system \eqref{eq:spform1} can be approximated by the blended dynamics with some kind of stability of the blended dynamics (and with sufficiently large $k$) as follows.

\begin{theorem}[\!\cite{kim2016robustness,lee2020tool}]\label{thm:cont_state}
Assume that the blended dynamics~\eqref{eq:bl1} is contractive.\footnote{$\dot x = f(t,x)$ is contractive if $\exists \Theta > 0$ such that $\Theta \frac{\partial f}{\partial x}(t,x) + \frac{\partial f}{\partial x}(t,x)^T \Theta \le - I$ for all $x$ and $t$ \cite{Slotine}.}
Then, for any compact set $K \subset \mathbb{R}^{nN}$ and for any $\eta > 0$, there exists $k^* > 0$ such that, for each $k > k^*$ and ${\rm col}(x_1(t_0), \dots, x_N(t_0)) \in K$, the solution to~\eqref{eq:net_state} exists for all $t \ge t_0$, and satisfies
\begin{align*}
    \limsup_{t \to \infty} \left\|x_i(t) - s(t)\right\| \le \eta, \quad \forall i \in \mathcal{N}.
\end{align*}
\end{theorem}

\begin{theorem}[\!\cite{panteley2017synchronization,lee2020tool}]\label{thm:set_state}
Assume that there is a nonempty compact set $\mathcal{A}_b \subset \mathbb{R}^n$ that is uniformly asymptotically stable for the blended dynamics~\eqref{eq:bl1}.
Let $\mathcal{D}_b \supset\mathcal{A}_b$ be an open subset of the domain of attraction of $\mathcal{A}_b$, and let\footnote{The condition for $w$ in ${\mathcal D}_x$ can be understood by recalling that ${\rm col}(x_1,\cdots,x_N) = 1_N \otimes s + (R \otimes I_n) \tilde z$.}
\begin{align*}
\mathcal{D}_x &:= \left\{1_N \otimes s + w: s \in \mathcal{D}_b, w \in \mathbb{R}^{nN} \,\, \text{such that} \,\, (1_N^T\otimes I_n)w = 0 \right\}.
\end{align*}
Then, for any compact set $K \subset \mathcal{D}_x\subset \mathbb{R}^{nN}$ and for any $\eta > 0$, there exists $k^* > 0$ such that, for each $k > k^*$ and ${\rm col}(x_1(t_0), \dots, x_N(t_0)) \in K$, the solution to~\eqref{eq:net_state} exists for all $t \ge t_0$, and satisfies
\begin{equation}\label{eq:as1}
\limsup_{t \to \infty} \|x_i(t)-x_j(t)\| \le \eta \quad \text{and} \quad \limsup_{t \to \infty} \|x_i(t)\|_{\mathcal{A}_b} \le \eta, \quad \forall i, j \in \NN.
\end{equation}
If, in addition, $\mathcal{A}_b$ is locally exponentially stable for the blended dynamics \eqref{eq:bl1} and, $f_i(t,s) = f_j(t,s)$, $\forall i, j \in \NN$, for each $s \in {\mathcal A}_b$ and $t$, then we have more than \eqref{eq:as1} as
$$\lim_{t \to \infty} \|x_i(t)-x_j(t)\| = 0 \quad \text{and} \quad \lim_{t \to \infty} \|x_i(t)\|_{\mathcal{A}_b} = 0, \quad \forall i, j \in \NN.$$
\end{theorem}

We emphasize the required stabilities in the above theorems are only for the blended dynamics \eqref{eq:bl1} but not for individual agent dynamics $\dot x_i = f_i(t,x_i)$.
A group of unstable and stable agents may end up with a stable blended dynamics, so that the above theorems can be applied.
In this case, it can be interpreted that the stability is traded throughout the network with strong couplings.

The blended dynamics \eqref{eq:bl1} shows an emergent behavior of the multi-agent system in the sense that $s(t)$ is governed by the new vector field that is assembled from the individual vector fields participating in the network.
From now, we list a few examples of designing multi-agent systems (or, simply called `networks') whose tasks are represented by the emergent behavior of \eqref{eq:bl1}, so that the whole network exhibits the emergent collective behavior with sufficiently large coupling gain $k$.

\subsection{Finding the number of agents participating in the network}

When constructing a distributed network, sometimes there is a need for each agent to know global information such as the number of agents in the network without resorting to a centralized unit.
In such circumstances, Theorem \ref{thm:cont_state} can be employed to design a distributed network that estimates the number of participating agents, under the assumption that there is one agent (whose index is 1 without loss of generality) who always takes part in the network.
Suppose that agent $1$ integrates the following scalar dynamics:
\begin{equation}\label{eq:N1}
\dot{x}_1 = -x_1 + 1 + k \sum_{j \in \NN_1} \alpha_{1j}(x_j - x_1)
\end{equation}
while all others integrate
\begin{equation}\label{eq:N2}
\dot{x}_i = 1 + k\sum_{j \in \NN_i} \alpha_{ij}(x_j - x_i), \quad i = 2, \dots, N
\end{equation}
where $N$ is unknown to the agents.
Then, the blended dynamics is obtained as
\begin{equation}\label{eq:Ns}
\dot{ s} = -\frac{1}{N}{s} + 1.
\end{equation}
This implies that the resulting emergent motion $s(t)$ converges to $N$ as time goes to infinity.
Then, it follows from Theorem \ref{thm:cont_state} that each state $x_i(t)$ approaches arbitrarily close to $N$ with a sufficiently large $k$.
Hence, by increasing $k$ such that the estimation error is less than $0.5$, and by rounding $x_i(t)$ to the nearest integer, each agent gets to know the number $N$ as time goes on.

By resorting to a heterogeneous network, we were able to impose stable emergent collective behavior that makes possible for individuals to estimate the number of agents in the network.
Note that the initial conditions do not affect the final value of $x_i(t)$ because they are forgotten as time tends to infinity due to the stability of the blended dynamics.
This is in sharp contrast to other approaches such as \cite{shames2012distributed} where the average consensus algorithm is employed, which yields the average of individual initial conditions, to estimate $N$.
While their approach requires resetting the initial conditions whenever some agents join or leave the network during the operation, the estimation of the proposed algorithm remains valid (after some transient) in such cases because the blended dynamics \eqref{eq:Ns} remains contractive for any $N \ge 1$.
Therefore, the proposed algorithm achieves the plug-and-play operation.
Moreover, when the maximum number $N_{\max}$ of agents is known, the decentralized design is also achieved.
Further details are found in \cite{donggil}.

\begin{remark}
	A slight variation of the idea yields an algorithm to identify the agents attending the network.
	Let the number $1$ in \eqref{eq:N1} and \eqref{eq:N2} be replaced by $2^{i-1}$, where $i$ is the unique ID of the agent in $\{1, 2, \dots, N_{\max}\}$.
	Then the blended dynamics \eqref{eq:Ns} becomes $\dot { s} = -(1/N)  s + \sum_{j \in \NN_a} 2^{j-1}/N$, where $\NN_a$ is the index set of the attending agents, and $N$ is the cardinality of $\NN_a$.
	Since the resulting emergent behavior $s(t) \to \sum_{j \in \NN_a} 2^{j-1}$, each agent can figure out the integer value $\sum_{j \in \NN_a} 2^{j-1}$, which contains the binary information of the attending agents.
\end{remark}

\subsection{Distributed least-squares solver}\label{subsec:lss}

Distributed algorithms have been developed in various fields of study so as to divide a large computational problem into small-scale computations. 
In this regard, finding a solution of a given large linear equation in a distributed manner has been tackled in recent years \cite{mou2013fixed, mou2015distributed, anderson2015decentralized}.
Let the equation be given by
\begin{align}\label{eq:le}
Ax =  b \quad \in \mathbb{R}^M
\end{align}
where $A \in \mathbb{R}^{M \times n}$ has full column rank, $x \in \mathbb{R}^n$, and $b \in \mathbb{R}^{M}$.
We suppose that the total of $M$ equations are grouped into $N$ equation banks, and the $i$-th equation bank consists of $m_i$ equations so that $\sum_{i=1}^N m_i = M$.
In particular, we write the $i$-th equation bank as
\begin{align}\label{eq:les}
A_ix = b_i \quad \in \mathbb{R}^{m_i}, \quad i = 1, 2, \dots, N,
\end{align}
where $A_i \in \mathbb{R}^{m_i \times n}$ is the $i$-th block rows of the matrix $A$, and $b_i \in \mathbb{R}^{m_i}$ is the $i$-th block elements of $b$.
The problem of finding a solution to \eqref{eq:les} (in the sense of least-squares when there is no solution) is dealt with in \cite{wang2017distributed, shi2016distributed, shi2017network, liu2017network}.
Most notable among them are \cite{shi2016distributed} and \cite{shi2017network}, in which they proposed a distributed algorithm given by
\begin{align}\label{eq:dls}
    \dot{x}_i = -A_i^T(A_ix_i - b_i) + k\sum_{j \in \mathcal{N}_i}\alpha_{ij}(x_j - x_i).
\end{align}

Here, we analyze \eqref{eq:dls} in terms of Theorem \ref{thm:set_state}.
In particular, the blended dynamics of the network~\eqref{eq:dls} is obtained as
\begin{align}\label{eq:lss}
    \dot{s} = -\frac{1}{N}A^T(As - b)
\end{align}
which is equivalent to the gradient descent algorithm of the optimization problem
\begin{align}
    \text{minimize}_x \,\, \|Ax - b\|^2
\end{align}
that has a unique minimizer $(A^TA)^{-1}A^Tb$; the least-squares solution of~\eqref{eq:le}.
Thus, Theorem \ref{thm:set_state} asserts that each state $x_i$ approximates the least-squares solution with a sufficiently large $k$, and the error can be made arbitrarily small by increasing $k$.

\begin{remark}
Even in the case that $A^TA$ is not invertible, the network \eqref{eq:dls} still solves the least-squares problem because $s$ of \eqref{eq:lss} converges to one of the minimizers.
Further details are found in \cite{JGLeeLCSS20}.
\end{remark}

\subsection{Distributed median solver}\label{subsec:med}

The idea of designing a network based on the gradient descent algorithm of an optimization problem, as in the previous subsection, can be used for most other distributed optimization problems.
Among them, a particularly interesting example is the problem of finding a median, which is useful, for instance, in rejecting outliers under redundancy.

For a collection $\mathcal{R}$ of real numbers $r_i$, $i=1, 2, \cdots, N$, their median is defined as a real number that belongs to the set
$${M}_{\mathcal{R}} = \begin{cases} \{r_{(N+1)/2}^s\}, &\mbox{ if $N$ is odd}, \\
[r_{N/2}^s, r_{N/2 + 1}^s], &\mbox{ if $N$ is even}, \end{cases}$$
where $r_i^s$'s are the elements of the set $\mathcal{R}$ with its index being rearranged (sorted) such that $r_1^s \le r_2^s \le \cdots \le r_N^s$.
With the help of this relaxed definition of the median, finding a median $s$ of ${\mathcal R}$ becomes solving an optimization problem
$$\text{minimize}_s \,\, \begin{matrix} \sum_{i=1}^N\end{matrix}|r_i - s|.$$
Then, a gradient descent algorithm given by
\begin{align}\label{eq:da_dms}
\dot{s} = \frac1N \sum_{i=1}^N  \text{sgn}(r_i - s)
\end{align}
will solve this minimization problem, where $\text{sgn}(s)$ is $1$ if $s > 0$, $-1$ if $s < 0$, and $0$ if $s = 0$.
In particular, the solution $s$ satisfies
$$\lim_{t \to \infty} \|s(t)\|_{{M}_{\mathcal{R}}} = 0.$$

Motivated by this, we propose a distributed median solver, whose individual dynamics of the agent $i$ uses the information of $r_i$ only: 
\begin{align}\label{eq:dms}
\dot{x}_i =  \, \text{sgn}(r_i - x_i) + k \sum_{j \in \mathcal{N}_i}\alpha_{ij} (x_j - x_i), \quad i \in \mathcal{N}.
\end{align}
Now, the algorithm \eqref{eq:dms} finds a median approximately by exchanging their states $x_i$ only (not $r_i$).
Further details can be found in~\cite{jgleeTAC19}.

\subsection{Distributed optimization: Optimal power dispatch}

As another application of distributed optimization, let us consider an optimization problem 
\begin{align}\label{eq:opt_prob}
\begin{split}
\text{minimize}_{\lambda_1,\cdots,\lambda_N} \quad &\sum_{i=1}^{N} J_i(\lambda_i) \\
\text{subject to} \quad &\sum_{i=1}^N \lambda_i = \sum_{i=1}^N d_i, \;\; \underline \lambda_i \le \lambda_i \le \overline \lambda_i, \;\; i \in \NN
\end{split}
\end{align}
where $\lambda_i \in \R$ is the decision variable, $J_i$ is a strictly convex $C^2$ function, and $\underline \lambda_i$, $\overline \lambda_i$, and $d_i$ are given constants.
A practical example is the economic dispatch problem of electric power, in which $d_i$ represents the demand of node $i$, $\lambda_i$ is the power generated at node $i$ with its minimum $\underline \lambda_i$ and maximum $\overline \lambda_i$, and $J_i$ is the generation cost.

A centralized solution is easily obtained using Lagrangian and Lagrange dual functions.
Indeed, it can be shown that the optimal value is obtained by $\lambda_i^* = \theta_i(s^*)$ where
$$\theta_i(s) := \left(\frac{dJ_i}{d\lambda_i}\right)^{-1}\left(\text{sat} \left(s, \frac{dJ_i}{d\lambda_i}(\underline{\lambda}_i), \frac{dJ_i}{d\lambda_i}(\overline{\lambda}_i)\right) \right),$$
in which $(dJ_i/d\lambda_i)^{-1}(\cdot)$ is the inverse function of $(dJ_i/d\lambda_i)(\cdot)$, $\text{sat}(s,a,b)$ is $s$ if $a \le s \le b$, $b$ if $b < s$, and $a$ if $s<a$.
The optimal $s^*$ maximizes the dual function $g(s) := \sum_{i=1}^N J_i(\theta_i(s)) + s (d_i - \theta_i(s))$, which is concave so that $s^*$ can be asymptotically obtained by the gradient algorithm:
\begin{equation}\label{eq:cen_edp}
\dot{s} = \frac{dg}{ds}(s) = \sum_{i=1}^N (d_i - \theta_i(s)).
\end{equation}

A distributed algorithm to solve the optimization problem approximately is to integrate
\begin{equation}\label{eq:do}
\dot{x}_i = d_i - \theta_i(x_i) + k \sum_{j \in \NN_i} \alpha_{ij} (x_j - x_i), \quad i\in \mathcal{N}
\end{equation}
because the blended dynamics of \eqref{eq:do} is given by
\begin{equation}\label{eq:bdo}
\dot{s} = \frac{1}{N} \sum_{i=1}^N (d_i - \theta_i(s)) = \frac1N \frac{dg}{ds}(s) .
\end{equation}
Obviously, \eqref{eq:bdo} is the same as the centralized solver \eqref{eq:cen_edp} except the scaling of $1/N$, which can be compensated by scaling \eqref{eq:do}.
By Theorem \ref{thm:set_state}, the state $x_i(t)$ of each node approaches arbitrarily close to $s^*$ with a sufficiently large $k$, and so, we obtain $\lambda_i^*$ approximately by $\theta_i(x_i(t))$ whose error can be made arbitrarily small.
Readers are referred to \cite{yun2018initialization}, which also describes the behavior of the proposed algorithm when the problem is infeasible so that each agent can figure out that infeasibility occurs.
It is again emphasized that the initial conditions are forgotten and so the plug-and-play operation is guaranteed.
Moreover, each agent can design its own dynamics \eqref{eq:do} only with their local information, so that decentralized design is achieved, except the global information $k$.
In particular, the function $\theta_i$ can be computed within the agent $i$ from the local information such as $J_i$, $\underline \lambda_i$, and $\overline \lambda_i$.
Therefore, the proposed solver \eqref{eq:do} does not exchange the private information of each agent (except the dual variable $x_i$).

\section{Strong diffusive output coupling}

Now, let us consider a bit more complex network---a heterogeneous multi-agent systems under diffusive output coupling law as\footnote{A particular case of \eqref{eq:net_output} is
\begin{equation*}
\dot x_i = f_i(t, x_i) +  k B \sum_{j \in \mathcal{N}_i}\alpha_{ij}(x_j - x_i), \quad i \in \mathcal{N},
\end{equation*}
where the matrix $B$ is positive semi-definite, which can always be converted into \eqref{eq:net_output} by a linear coordinate change.}
\begin{align}\label{eq:net_output}
\begin{split}
\dot{z}_i &= g_i(t, z_i, y_i) \;\quad\quad\quad\quad\quad\quad\quad\quad\quad \in \mathbb{R}^{m_i}, \\
\dot{y}_i &= h_i(t, y_i, z_i) +  k\Lambda\sum_{j \in \mathcal{N}_i}\alpha_{ij}(y_j - y_i)\;\;\in \mathbb{R}^n, \qquad i \in \NN,
\end{split}
\end{align}
where the matrix $\Lambda$ is positive definite.
The vector fields $g_i$ and $h_i$ are assumed to be piecewise continuous in $t$, continuously differentiable with respect to $z_i$ and $y_i$, locally Lipschitz with respect to $z_i$ and $y_i$ uniformly in $t$, and $g_i(t, 0, 0)$, $h_i(t, 0, 0)$ are uniformly bounded for $t$.

For this network, under the same coordinate change as \eqref{eq:ct1} in which $x_i$ replaced by $y_i$, it can be seen that the quasi-steady-state subsystem (or, the blended dynamics) becomes
\begin{align}\label{eq:blend_output}
\begin{split}
\dot{\hat{z}}_i &= g_i(t, \hat{z}_i, s), \quad i \in \mathcal{N}, \\
\dot{s} &= \frac{1}{N}\sum_{i=1}^N h_i(t, s, \hat{z}_i).
\end{split}
\end{align}
This can also be seen by treating $z_i(t)$ as external inputs of $h_i$ in \eqref{eq:net_output}.

\begin{theorem}[\!\cite{lee2020tool}]\label{thm:cont_output}
Assume that the blended dynamics \eqref{eq:blend_output} is contractive.
Then, for any compact set $K$ and for any $\eta > 0$, there exists $k^* > 0$ such that, for each $k > k^*$ and ${\rm col}(z_1(t_0),y_1(t_0),\dots,z_N(t_0),y_N(t_0)) \in K$, the solution to \eqref{eq:net_output} exists for all $t \ge t_0$, and satisfies
\begin{align*}
\limsup_{t \to \infty} \|z_i(t) - \hat z_i(t)\| \le \eta \quad \text{and} \quad \limsup_{t \to \infty} \|y_i(t) - s(t)\| \le \eta, \quad \forall i \in \NN.
\end{align*}
\end{theorem}

\begin{theorem}[\!\cite{panteley2017synchronization,lee2020tool}]\label{thm:set_output}
Assume that there is a nonempty compact set $\mathcal{A}_b$ that is uniformly asymptotically stable for the blended dynamics \eqref{eq:blend_output}.
Let $\mathcal{D}_b \supset\mathcal{A}_b$ be an open subset of the domain of attraction of $\mathcal{A}_b$, and let
\begin{align*}
\mathcal{A}_x &:= \left\{  {\rm col}(\hat{z}_1, s, \hat{z}_2, s, \dots, \hat{z}_N, s) : {\rm col}(\hat{z}_1, \dots, \hat{z}_N, s) \in \mathcal{A}_b\right\}, \\
\mathcal{D}_x &:= \left\{ {\rm col}(\hat{z}_1, s_1, \dots, \hat{z}_N, s_N) : {\rm col}(\hat{z}_1, \dots, \hat{z}_N, s) \in \mathcal{D}_b \,\, \text{such that} \,\, \frac1N\sum_{i=1}^Ns_i = s\right\}.
\end{align*}
Then, for any compact set $K \subset \mathcal{D}_x$ and for any $\eta > 0$, there exists $k^* > 0$ such that, for each $k > k^*$ and ${\rm col}(z_1(t_0),y_1(t_0),\dots,z_N(t_0),y_N(t_0)) \in K$, the solution to \eqref{eq:net_output} exists for all $t \ge t_0$, and satisfies
\begin{equation}\label{eq:as2}
\limsup_{t \to \infty}\|{\rm col}(z_1(t),y_1(t),\dots,z_N(t),y_N(t))\|_{\mathcal{A}_x} \le \eta.
\end{equation}
If, in addition, $\mathcal{A}_b$ is locally exponentially stable for the blended dynamics \eqref{eq:blend_output} and, $h_i(t,y_i,z_i) = h_j(t,y_j,z_j)$, $\forall i, j \in \NN$, for each ${\rm col}(z_1,y_1,\dots,z_N,y_N) \in {\mathcal A}_x$ and $t$, then we have more than \eqref{eq:as2} as
$$\lim_{t\to\infty}\|{\rm col}(z_1(t),y_1(t),\dots,z_N(t),y_N(t))\|_{\mathcal{A}_x} = 0.$$
\end{theorem}

With the extended results, two more examples follow.

\subsection{Synchronization of heterogeneous Li{\'e}nard systems}\label{subsec:lie}

Consider a network of heterogeneous Li{\'e}nard systems modeled as
\begin{align}\label{eq:eachlie1}
\ddot z_i + f_i(z_i) \dot z_i + g_i(z_i) = u_i, \qquad i = 1, \cdots, N,
\end{align}
where $f_i(\cdot)$ and $g_i(\cdot)$ are locally Lipschitz.
Suppose that the output and the diffusive coupling input are given by
\begin{align}\label{eq:eachlie2}
o_i = a z_i + \dot z_i, \quad a>0, \quad \text{and} \quad u_i = k \sum_{j \in \mathcal{N}_i} \alpha_{ij} (o_j - o_i).
\end{align}
For \eqref{eq:eachlie1} with \eqref{eq:eachlie2}, we claim that synchronous and oscillatory behavior is obtained with a sufficiently large $k$ if the {\em averaged Li{\'e}nard systems} given by
\begin{align}\label{eq:liecondi}
\ddot z + \hat{f}(z) \dot z + \hat{g}(z) := \ddot z + \left(\frac{1}{N}\sum_{i=1}^N f_i(z)\right) \dot z + \left(\frac{1}{N}\sum_{i=1}^N g_i(z)\right) = 0
\end{align}
has a stable limit cycle.
This condition may be interpreted as the blended version of the condition for a stand-alone Li{\'e}nard system $\ddot z + f(z)\dot z + g(z) = 0$ to have a stable limit cycle.
Note that this condition implies that, even when some particular agents $\ddot z_i + f_i(z_i)\dot z_i + g_i(z_i) = 0$ do not yield a stable limit cycle, the network still can exhibit oscillatory behavior as long as the average of $(f_i,g_i)$ yields a stable limit cycle.
It is seen that stability of individual agents can be traded among agents in this way, so that some malfunctioning oscillators can oscillate in the oscillating network as long as there are a majority of good neighbors.
The frequency and the shape of synchronous oscillation is also determined by the average of $(f_i, g_i)$.

To justify the claim, we first realize \eqref{eq:eachlie1} and \eqref{eq:eachlie2} with $y_i := a z_i + \dot z_i$ as
\begin{align*}
\dot z_i &= - a z_i + y_i \\
\dot y_i &= -a^2 z_i + a y_i - f_i(z_i) y_i + a f_i(z_i) z_i - g_i(z_i) + k \sum_{j \in \mathcal{N}_i} \alpha_{ij}(y_j - y_i),
\end{align*}
and compute its blended dynamics \eqref{eq:blend_output} as
\begin{align}\label{eq:lieqss}
\dot{\hat{z}}_i &= - a \hat{z}_i + s, \qquad i \in \mathcal{N}, \\
\dot s &= - a^2 \left(\frac{1}{N} \sum_{i=1}^N \hat{z}_i\right) + as - \left(\frac{1}{N}\sum_{i=1}^N f_i(\hat{z}_i)\right) s + a \left(\frac{1}{N}\sum_{i=1}^N f_i(\hat{z}_i) \hat{z}_i\right)  - \left(\frac{1}{N}\sum_{i=1}^N g_i(\hat{z}_i)\right). \nonumber
\end{align}
To see whether this $(N+1)$-th order blended dynamics has a stable limit cycle, we observe that, with $a>0$, all $\hat z_i(t)$ converge exponentially to a common trajectory $\hat z(t)$ as time goes on.
Therefore, if the blended dynamics has a stable limit cycle, which is an invariant set, it has to be on the synchronization manifold $\mathcal{S}$ defined as
$$\mathcal{S} := \left\{ {\rm col}(\hat{z}, \dots, \hat{z}, \bar{s}) \in \mathbb{R}^{N+1} : {\rm col}(\hat{z}, \bar{s}) \in \mathbb{R}^2\right\}.$$
Projecting the blended dynamics \eqref{eq:lieqss} to the synchronization manifold $\mathcal{S}$, i.e., replacing $\hat{z}_i$ with $\hat z$ in \eqref{eq:lieqss} for all $i \in \mathcal{N}$, we obtain a second-order system
\begin{align}\label{eq:zs}
\begin{split}
\dot {\hat z} &= -a \hat z + s, \\
\dot s &= -a^2 \hat{z} + as - \hat{f}(\hat{z}) s + a \hat f(\hat z) \hat z - \hat{g}(\hat{z}).
\end{split}
\end{align}
Therefore, \eqref{eq:zs} should have a stable limit cycle if the blended dynamics has a stable limit cycle.
It turns out that \eqref{eq:zs} is a realization of \eqref{eq:liecondi} by $s = az + \dot z$, and thus, existence of a stable limit cycle for \eqref{eq:liecondi} is a necessary condition for the blended dynamics \eqref{eq:lieqss} to have a stable limit cycle.
Further analysis, given in \cite{jingyuNOLCOS}, proves that the converse is also true.
Then, Theorem \ref{thm:set_output} holds with the limit cycle of \eqref{eq:lieqss} as the compact set ${\mathcal A}_b$, and thus, with a sufficiently large $k$, all the vectors $(z_i(t),\dot z_i(t))$ stay close to each other, and oscillate near the limit cycle of the averaged Li{\'e}nard system \eqref{eq:liecondi}.
This property has been coined as `phase cohesiveness' in \cite{Dorfler14}.

\subsection{Distributed state estimation}\label{subsec:de}

Consider a linear system
$$\dot{\chi} = S\chi \; \in \R^n, \quad 
o = \begin{bmatrix} o_1 \\ \vdots \\ o_N \end{bmatrix} = \begin{bmatrix} G_1 \\ \vdots \\ G_N \end{bmatrix} \chi + \begin{bmatrix} n_1 \\ \vdots \\ n_N\end{bmatrix} = G \chi + n, \quad 
o_i \in \R^{q_i}$$
where $\chi \in \R^n$ is the state to be estimated, $o$ is the measurement output, and $n$ is the measurement noise.
It is supposed that there are $N$ distributed agents, and each agent $i$ can access the measurement $o_i \in \R^{q_i}$ only (where often $q_i=1$).
We assume that the pair $(G,S)$ is detectable, while each pair $(G_i,S)$ is not necessarily detectable as in \cite{bai2011distributed,kim2016distributed}.
Each agent is allowed to communicate its internal state to its neighboring nodes.
The question is how to construct a dynamic system for each node that estimates $\chi(t)$.
See, e.g., \cite{mitra2016approach,kim2016distributedlue} for more details on this distributed state estimation problem.

To solve the problem, we first employ the detectability decomposition for each node, that is, for each pair $(G_i,S)$.
With $p_i$ being the dimension of the undetectable subspace of the pair $(G_i,S)$, let $[Z_i, W_i]$ be an orthogonal matrix, where $Z_i \in \R^{n \times (n-p_i)}$ and $W_i \in \R^{ n \times p_i}$, such that
$$\begin{bmatrix} Z_i^T \\ W_i^T \end{bmatrix} S \begin{bmatrix}Z_i & W_i\end{bmatrix}
= \begin{bmatrix} \bar{S}_i & 0 \\ * & * \end{bmatrix}, \quad 
G_i  \begin{bmatrix} Z_i & W_i\end{bmatrix}= \begin{bmatrix}\bar{G}_i & 0\end{bmatrix}$$
and the pair $(\bar{G}_i, \bar{S}_i)$ is detectable.
Then, pick a matrix $\bar U_i \in \R^{(n-p_i) \times q_i}$ such that $\bar S_i - \bar U_i \bar G_i$ is Hurwitz.
Now, individual agent $i$ can construct a local partial state observer, for instance, as
\begin{align}\label{eq:part_obs}
\dot{b}_i = \bar{S}_i b_i - \bar{U}_i(\bar{G}_ib_i - o_i) \quad \in \mathbb{R}^{p_i}
\end{align}
to collect the information of the state $\chi$ as much as possible from the available measurement $o_i$ only; each agent $i$ can obtain a partial information $b_i$ about $\chi$ in the sense that
\begin{align}\label{eq:le_de}
b_i = Z_i^T \chi + z_i \quad \in \mathbb{R}^{p_i}, \quad i\in \mathcal{N}
\end{align}
where $z_i$ denotes the estimation error that converges to zero by \eqref{eq:part_obs} if $n_i=0$.
When we collect \eqref{eq:le_de} and write them as 
\begin{align}\label{eq:le_des}
b = \begin{bmatrix} b_1 \\ \vdots \\ b_N \end{bmatrix} = \begin{bmatrix} Z_1^T \\ \vdots \\ Z_N^T \end{bmatrix} \chi + \begin{bmatrix} z_1 \\ \vdots \\ z_N \end{bmatrix} =: A \chi + z,
\end{align}
detectability of $(G, S)$ implies that ${\rm col}(Z_1^T, \dots, Z_N^T) = A$ has full-column rank.
Therefore, the least-squares solution $\hat \chi(t)$ of $A \hat \chi(t) = b(t)$ can generate a unique estimate of $\chi(t)$.
This reminds us of the problem in Section~\ref{subsec:lss}; finding the least-squares solution in a distributed manner.

Based on the discussion above, we propose a distributed state estimator for the given linear system as
\begin{align}\label{eq:de_net}
\dot{\hat{\chi}}_i &= S\hat{\chi}_i - \kappa Z_i(Z_i^T \hat{\chi}_i - b_i) + k \sum_{j \in \mathcal{N}_i} \alpha_{ij} (\hat{\chi}_j - \hat{\chi}_i) \quad \in \R^n, \quad i\in \mathcal{N},
\end{align}
where $b_i$ comes from \eqref{eq:part_obs}, and both $\kappa$ and $k$ are design parameters.
Note that the least-squares solution $\hat \chi$ for $A\hat \chi(t) = b(t)$ is time-varying, and so, in order to have asymptotic convergence of $\hat \chi_i(t)$ to $\chi(t)$ (when there is no noise $n$), we had to add the generating model of $\chi(t)$ in \eqref{eq:de_net}, inspired by the internal model principle.

To justisfy the proposed distributed state estimator \eqref{eq:part_obs} and \eqref{eq:de_net}, let us denote the state estimation error as $y_i := \hat \chi_i - \chi$, then we obtain the error dynamics for the partial state observer~\eqref{eq:part_obs} and the distributed observer~\eqref{eq:de_net} as
\begin{align}\label{eq:de_error}
\begin{split}
\dot{z}_i &= (\bar{S}_i - \bar{U}_i\bar{G}_i) z_i + \bar{U}_i n_i \\
\dot{y}_i &= S y_i - \kappa Z_i(Z_i^T y_i - z_i) + k \sum_{j \in \mathcal{N}_i} \alpha_{ij}(y_j - y_i), \qquad i \in \NN.
\end{split}
\end{align}
The blended dynamics \eqref{eq:blend_output} is obtained as
\begin{align}\label{eq:blend_de_lss}
\dot{\hat{z}}_i &= (\bar{S}_i - \bar{U}_i\bar{G}_i)\hat{z}_i + \bar{U}_in_i \nonumber \\
\dot{s} &= Ss - \frac{\kappa}{N}\sum_{i=1}^N Z_i(Z_i^T s - \hat{z}_i) = \left( S - \frac{\kappa}{N}A^T A \right) s + \frac{\kappa}{N}A^T \begin{bmatrix} \hat{z}_1 \\ \vdots \\ \hat{z}_N\end{bmatrix}.
\end{align}
For a sufficiently large gain $\kappa$, the blended dynamics~\eqref{eq:blend_de_lss} becomes contractive, and thus, Theorem \ref{thm:cont_output} guarantees that the error variables $(z_i(t),y_i(t))$ of \eqref{eq:de_error} behave like $(\hat z_i(t),s(t))$ of \eqref{eq:blend_de_lss}.
Moreover, if there is no noise $n$, Theorem \ref{thm:set_output} asserts that all the estimation errors $(z_i(t),y_i(t))$, $i \in \NN$, converge to zero because ${\mathcal A}_x = \{0\}$.
Even with the noise $n$, the proposed observer achieves almost best possible estimate whose detail can be found in \cite{JGLeeLCSS20}.

\section{General description of the blended dynamics}

Now, we extend our approach to the most general setting---a heterogeneous multi-agent systems under rank-deficient diffusive coupling law given by
\begin{align}\label{eq:net_rank}
\dot x_i = f_i(t, x_i) + k B_i \sum_{j \in \mathcal{N}_i} \alpha_{ij} \left(x_j-x_i\right), \quad i \in \mathcal{N},
\end{align}
where the matrix $B_i$ is positive semi-definite for each $i \in \NN$.
For this network, by increasing the coupling gain $k$, we can enforce synchronization of the states that correspond to the subspace
\begin{equation}\label{eq:rb}
R_B := \bigcap_{i=1}^N \text{im}(B_i) \quad \subset \R^n.
\end{equation}
In order to find the part of individual states that synchronize, let us follow the procedure:
\begin{enumerate}

\item Find $W_i \in \mathbb{R}^{n \times p_i }$ and $Z_i \in \mathbb{R}^{n \times (n - p_i)}$ where $p_i$ is the rank of $B_i$, such that $[W_i \; Z_i]$ is an orthogonal matrix and
\begin{align}
\begin{bmatrix} W_i^T \\ Z_i^T \end{bmatrix} B_i \begin{bmatrix} W_i & Z_i \end{bmatrix} =  \begin{bmatrix} \Lambda_i^2 & 0 \\ 0 & 0 \end{bmatrix} \label{eq:Bi}
\end{align}
where $\Lambda_i \in \mathbb{R}^{p_i \times p_i}$ is positive definite.
Let ${W}_\text{net} := {\rm diag}(W_1, \dots, W_N)$, ${Z}_\text{net} := {\rm diag}(Z_1, \dots, Z_N)$, and ${\Lambda}_\text{net} := {\rm diag}(\Lambda_1, \dots, \Lambda_N)$.

\item Find $V_i \in \R^{p_i \times p_s}$ such that, with $\bar p := \sum_{i=1}^N p_i$ and $V := {\rm col}(V_1, \dots, V_N) \in \R^{\bar p \times p_s}$, the columns of $V$ are orthonormal vectors satisfying
\begin{equation}\label{eq:TInull}
(\mathcal{L} \otimes I_n) {W}_\text{net} {\Lambda}_\text{net} V = 0_{n(N-1) \times p_s}
\end{equation}
where $p_s$ is the dimension of ${\rm ker}(\mathcal{L} \otimes I_n) {W}_\text{net} {\Lambda}_\text{net}$, and $\LL$ is the graph Laplacian matrix.

\item Find $\overline V \in \R^{\bar p \times (\bar p-p_s)}$ such that $[V \; \overline V] \in \R^{\bar p \times \bar p}$ is an orthogonal matrix.

\end{enumerate}

\begin{proposition}[\!\cite{lee2020tool}]
\begin{enumerate}
\item [(i)] $p_s \le \min\{p_1,\dots,p_N\} \le n$. 
\item [(ii)] All matrices $W_i \Lambda_i V_i$ ($i=1,\cdots,N$) are the same; so let us denote it by $M \in \R^{n \times p_s}$, then ${\rm rank}(M) = p_s$, ${\rm im}(M) = R_B$, and $\dim(R_B) = p_s$.
\item [(iii)] Define
$$Q := \overline V^T {\Lambda}_{\rm net} {W}_{\rm net}^T (\mathcal{L} \otimes I_n) {W}_{\rm net} {\Lambda}_{\rm net} \overline V \quad \in \R^{(\bar p-p_s) \times (\bar p-p_s)}.$$
Then, $Q$ is positive definite.
\end{enumerate}
\end{proposition}

Now, we introduce a linear coordinate change by which the state $x_i$ of individual agent is split into $Z_i^T x_i$ and $W_i^T x_i$.
In particular, the sub-state $z_i := Z_i^Tx_i$ is the projected component of $x_i$ on ${\rm im} (Z_i)$, and has no direct interconnection with the neighbors as its dynamics is given by
\begin{equation}\label{eq:zi}
\dot{z}_i = Z_i^Tf_i(t, x_i), \qquad i \in \NN.
\end{equation}
On the other hand, the sub-state $W_i^Tx_i$ is split once more into $s_i := V_i^T \Lambda_i^{-1} W_i^T x_i \in \R^{p_s}$ and the other part.
(In fact, $s_i$ determines the behavior of the individual agent in the subspace $R_B$ in the sense that $M s_i \in R_B$.)
With a sufficiently large $k$, these $s_i$ are enforced to synchronize to $s := (1/N)\sum_{i=1}^N s_i = (1/N)\sum_{i=1}^N V_i^T \Lambda_i^{-1} W_i^T x_i$, which is governed by
\begin{equation}\label{eq:s}
\dot s = \frac1N \sum_{i=1}^N V_i^T\Lambda_i^{-1} W_i^T f_i(t,x_i).
\end{equation}
To see this, let us consider a coordinate change for the whole multi-agent system \eqref{eq:net_rank}:
\begin{align}\label{eq:state_transformation_rank}
\begin{bmatrix} z \\ s \\ w \end{bmatrix} = \begin{bmatrix} Z_\text{net}^T \\ (1/N)V^T\Lambda_\text{net}^{-1}W_\text{net}^T \\ Q^{-1}\overline{V}^T\Lambda_\text{net}W_\text{net}^T(\mathcal{L} \otimes I_n)\end{bmatrix} \begin{bmatrix} x_1 \\ \vdots \\ x_N\end{bmatrix}  
\end{align}
where $w \in \R^{(N-1)p_s + \sum_{i=1}^N(p_i-p_s)}$ collects all the components both in ${\rm col}(s_1,\dots,s_N)$ that are left after taking $s = (1/N) 1_N^T {\rm col}(s_1,\dots,s_N) \in \R^{p_s}$, and in $W_i^Tx_i$ that are left after taking $s_i = V_i^T \Lambda_i^{-1} W_i^T x_i$.
It turns out that the governing equation for $w$ is
\begin{equation}\label{eq:w}
\frac1k \dot w =  - Q w + \frac1k Q^{-1} \overline{V}^T \Lambda_\text{net} W_\text{net}^T (\mathcal{L} \otimes I_n) \begin{bmatrix} f_1(t, x_1) \\ \vdots \\ f_N(t, x_N) \end{bmatrix}.
\end{equation}
Then, it is clear that the system \eqref{eq:zi}, \eqref{eq:s}, and \eqref{eq:w} is in the standard form of singular perturbation.
Since the inverse of \eqref{eq:state_transformation_rank} is given (in \cite{lee2020tool}) by
\begin{align*}
\begin{bmatrix} x_1 \\ \vdots \\ x_N \end{bmatrix} = (Z_\text{net} - W_\text{net}\Lambda_\text{net}L)z + NW_\text{net}\Lambda_\text{net}Vs + W_\text{net}\Lambda_\text{net}\overline{V}w
\end{align*}
where $L \in \mathbb{R}^{\bar p \times (nN - \bar p)}$ is defined as
\begin{equation*}
L = {\rm col}( L_1, \dots,  L_N ):= \overline V Q^{-1} \overline V^T {\Lambda}_\text{net} {W}_\text{net}^T (\mathcal{L} \otimes I_n) {Z}_\text{net}
\end{equation*}
with $L_i \in \mathbb{R}^{p_i \times (nN - \bar p)}$, the quasi-steady-state subsystem (that is, the blended dynamics) becomes
\begin{align}\label{eq:blend_rank}
\begin{split}
\dot{\hat{z}}_i &= Z_i^T f_i(t, Z_i\hat{z}_i - W_i\Lambda_i L_i \hat{z} + NMs), \quad i \in \mathcal{N} \\
\dot{s} &= \frac{1}{N}\sum_{i=1}^N V_i^T\Lambda_i^{-1}W_i^T f_i(t, Z_i\hat{z}_i - W_i\Lambda_i L_i \hat{z} + NMs)
\end{split}
\end{align}
where $\hat z = {\rm col}(\hat z_1,\dots,\hat z_N)$.

\begin{theorem}[\!\cite{lee2020tool}]\label{thm:cont_rank}
Assume that the blended dynamics~\eqref{eq:blend_rank} is contractive.
Then, for any compact set $K$ and for any $\eta > 0$, there exists $k^* > 0$ such that, for each $k > k^*$ and ${\rm col}(x_1(t_0),\dots,x_N(t_0)) \in K$, the solution to \eqref{eq:net_rank} exists for all $t \ge t_0$, and satisfies
\begin{align*}
\limsup_{t \to \infty} \left\| x_i(t) - (Z_i \hat{z}_i(t) - W_i\Lambda_iL_i\hat{z}(t) + NMs(t)) \right\| \le \eta, \qquad \forall i \in \NN.
\end{align*}
\end{theorem}

\begin{theorem}[\!\cite{lee2020tool}]\label{thm:set_rank}
Assume that there is a nonempty compact set $\mathcal{A}_b$ that is uniformly asymptotically stable for the blended dynamics \eqref{eq:blend_rank}.
Let $\mathcal{D}_b \supset\mathcal{A}_b$ be an open subset of the domain of attraction of $\mathcal{A}_b$, and let
\begin{align*}
\mathcal{A}_x &:= \left\{ (Z_{\rm net} - W_{\rm net}\Lambda_{\rm net}L) \hat{z} + N(1_N \otimes M) s : {\rm col}(\hat{z}, s) \in \mathcal{A}_b \right\}, \\
\mathcal{D}_x &:= \left\{(Z_{\rm net} - W_{\rm net}\Lambda_{\rm net}L)\hat{z} + N(1_N \otimes M) s + W_{\rm net}\Lambda_{\rm net}\overline{V} w : \begin{bmatrix}\hat{z}\\ s\end{bmatrix} \in \mathcal{D}_b, w \in \mathbb{R}^{\bar{p} - p_s}\right\}.
\end{align*}
Then, for any compact set $K \subset \mathcal{D}_x$ and for any $\eta > 0$, there exists $k^* > 0$ such that, for each $k > k^*$ and ${\rm col}(x_1(t_0), \dots, x_N(t_0)) \in K$, the solution to \eqref{eq:net_rank} exists for all $t \ge t_0$, and satisfies
\begin{align}\label{eq:thm6}
\limsup_{t \to \infty} \left \|{\rm col}(x_1(t), \dots, x_N(t))\right\|_{\mathcal{A}_x} \le \eta.
\end{align}
If, in addition, $\mathcal{A}_b$ is locally exponentially stable for the blended dynamics \eqref{eq:blend_rank} and, 
\begin{align}\label{eq:thm6-1}
\overline{V}^T\Lambda_{\rm net}W_{\rm net}^T(\mathcal{L}\otimes I_n)\begin{bmatrix} f_1(t, x_1) \\ \vdots \\ f_N(t, x_N) \end{bmatrix} = 0,
\end{align}
for all ${\rm col}(x_1,\dots,x_N) \in \mathcal{A}_x$ and $t$, then we have more than \eqref{eq:thm6} as
$$\lim_{t\to\infty}\left\|{\rm col}(x_1(t), \dots, x_N(t))\right\|_{\mathcal{A}_x} = 0.$$
\end{theorem}

\subsection{Distributed state observer with rank-deficient coupling}

We revisit the distributed state estimation problem discussed in Section~\ref{subsec:de} with the following agent dynamics, which has less dimension than \eqref{eq:part_obs} and \eqref{eq:de_net}:
\begin{align}\label{eq:dist_state_obs}
\dot{\hat{\chi}}_i = S \hat{\chi}_i + U_i (o_i - G_i\hat{\chi}_i) + k W_iW_i^T \sum_{j=1}^N \alpha_{ij} (\hat \chi_j - \hat \chi_i)
\end{align}
where $U_i := Z_i\bar{U}_i$ and $k$ is sufficiently large.
Here, the first two terms on the right-hand side look like a typical state observer, but due to the lack of detectability of $(G_i,S)$, it cannot yield stable error dynamics.
Therefore, the diffusive coupling of the third term exchanges the internal state with the neighbors, compensating for the lack of information on the undetectable parts.
Recalling that $W_i^T\chi$ represents the undetectable components of $\chi$ by $o_i$ in the decomposition given in Section~\ref{subsec:de}, it is noted that the coupling term compensates only the undetectable portion in the observer.
As a result, the coupling matrix $W_i W_i^T$ is rank-deficient in general.
This point is in sharp contrast to the previous results such as \cite{kim2016distributedlue}, where the coupling term is nonsingular so that the design is more complicated.

With $x_i := \hat \chi_i - \chi$ and $B_i = W_i W_i^T$, the error dynamics becomes
$$\dot x_i = (S - U_iG_i)x_i + U_i n_i + k B_i \sum_{j=1}^N \alpha_{ij} (x_j-x_i), \qquad i \in \NN.$$
This is precisely the multi-agent system \eqref{eq:net_rank}, where in this case the matrices $Z_i$ and $W_i$ have implications related to detectable decomposition.
In particular, from the detectability of the pair $(G,S)$, it is seen that $\cap_{i=1}^N {\rm im} (W_i) = \cap_{i=1}^N \ker (Z_i^T) = \{0\}$ (which corresponds to $R_B$ in \eqref{eq:rb}), by the fact that $\ker (Z_i^T)$ is the undetectable subspace of the pair $(G_i, S)$.
This implies that $p_s=0$, $V$ is null, and thus, $\overline{V}$ can be chosen as the identity matrix.
With them, the blended dynamics \eqref{eq:blend_rank} is given by, with the state $s$ being null,
\begin{align*}
\dot{\hat{z}}_i &= Z_i^T (S-U_iG_i) (Z_i\hat{z}_i - W_i\Lambda_i L_i \hat{z}) + Z_i^T U_i n_i \\
&= (Z_i^T S - \bar U_i \bar G_i Z_i^T) (Z_i \hat z_i - W_i \Lambda_i L_i \hat z) + \bar U_i n_i = (\bar S_i - \bar U_i \bar G_i) \hat z_i + \bar U_i n_i, \quad i \in \NN.
\end{align*}
Since $\bar{S}_i - \bar{U}_i\bar{G}_i$ is Hurwitz for all $i$, the blended dynamics is contractive and Theorem~\ref{thm:cont_rank} asserts that the estimation error $x_i(t)$ behaves like $Z_i\hat{z}_i(t)$, with a sufficiently large $k$.
Moreover, if there is no measurement noise, then the set $\AAA_b = \{0\} \subset \R^{nN-\bar p}$ is globally exponentially stable for the blended dynamics.
Then, Theorem \ref{thm:set_rank} asserts that the proposed distributed state observer exponentially finds the correct estimate with a sufficiently large $k$ because \eqref{eq:thm6-1} holds with $\AAA_x = \{0\} \subset \R^{nN}$.

\section{Robustness of emergent collective behavior}

When a product is manufactured in a factory, or some cells and organs are produced in an organism, a certain level of variance is inevitable due to imperfection of the production process.
In this case, how to reduce the variance in the outcomes if improving the process itself is not easy or even impossible?

We have seen throughout the chapter, the emergent collective behavior of the network involves averaging the vector fields of individual agents; that is, the network behavior is governed by the blended dynamics if the coupling strength is sufficiently large.
Therefore, even if the individual agents are created with relatively large variance from their reference model, its blended dynamics can have smaller variance because of the averaging effect.
Then, when the coupling gain is large, all the agents, which were created with large variance, can behave like an agent that is created with small variance.

In this section, we illustrate this point.
In particular, we simulate a network of pacemaker cells under a single conduction line.
The nominal behavior of a pacemaker cell is given in \cite{dos2004rhythm} as
\begin{equation}\label{eq:pm}
\ddot z + (1.45 z^2 - 2.465 z - 0.551) \dot z + z = 0
\end{equation}
which is a Li{\'e}nard system considered in Section \ref{subsec:lie} that has a stable limit cycle.
Now, suppose that a group of pacemaker cells are produced with some uncertainty so that they are represented as
$$\ddot z_i + f_i(z_i) \dot{z}_i + g_i(z_i) = u_i, \qquad i = 1, \dots, N,$$
where
\begin{align*}
f_i(z_i) &= 0.1\Delta_i^1 z_i^3 + (1.45+\Delta_i^2) z_i^2 - (2.465+\Delta_i^3) z_i - (0.551+\Delta_i^4) \\
g_i(z_i) &= (1+\Delta_i^5) z_i + 0.1\Delta_i^6 z_i^2
\end{align*}
in which, all $\Delta_i^l$ are randomly chosen from a distribution of zero mean and unit variance.
With $u_i = k \sum_{j \in \NN_i} (\dot z_j + z_j - \dot z_i - z_i)$, the blended dynamics of the group of pacemaker is given as the averaged Li{\'e}nard system \eqref{eq:liecondi} with
\begin{align*}
\hat f(z) &= 0.1 \bar \Delta^1 z^3 + (1.45+\bar\Delta^2) z^2 - (2.465+\bar\Delta^3) z - (0.551+\bar\Delta^4) \\
\hat g(z) &= (1+\bar\Delta^5) z + 0.1\bar\Delta^6 z^2
\end{align*}
where $\bar\Delta^l = (1/N)\sum_{i=1}^N \Delta_i^l$ whose expectation is zero and variance is $1/N$.
By the Chebyshev's theorem in probability, it is seen that the behavior of the blended dynamics recovers that of \eqref{eq:pm} almost surely as $N$ tends to infinity.
It is emphasized that some agent may not have a stable limit cycle depending on their random selection of $\Delta_i^l$, but the network can still exhibit oscillatory behavior, and the frequency and the shape of oscillation becomes more robust as the number of agents gets large.

\begin{figure}
\centering
\subfloat[$N=10$]{\includegraphics[width=.32\textwidth]{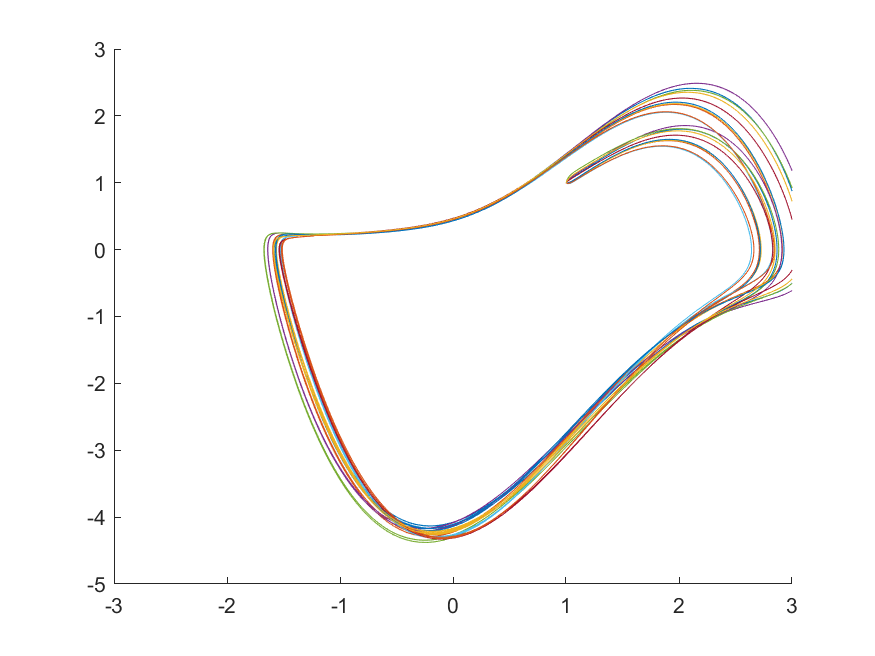}}
\subfloat[$N=10$]{\includegraphics[width=.32\textwidth]{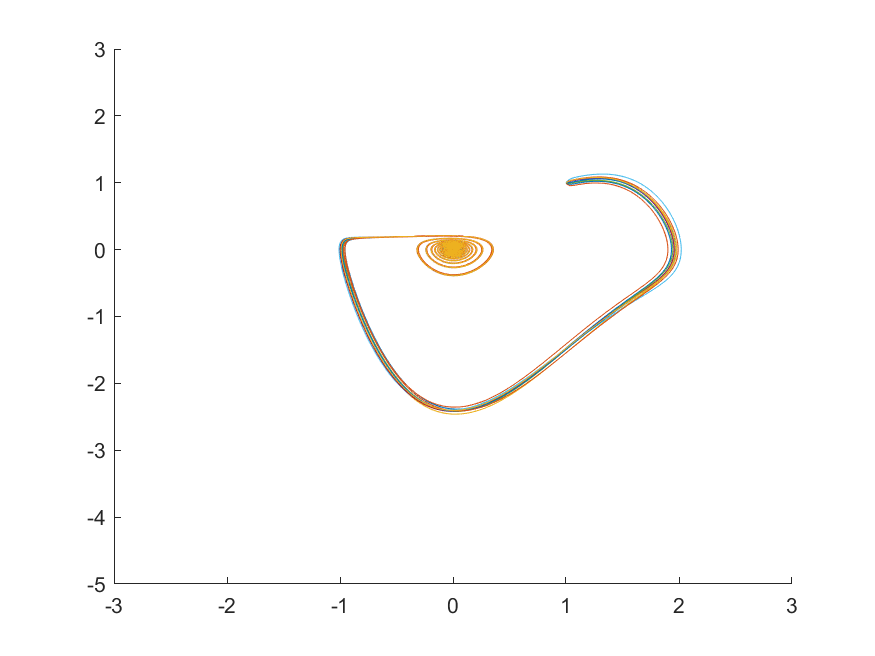}}
\subfloat[$N=10$]{\includegraphics[width=.32\textwidth]{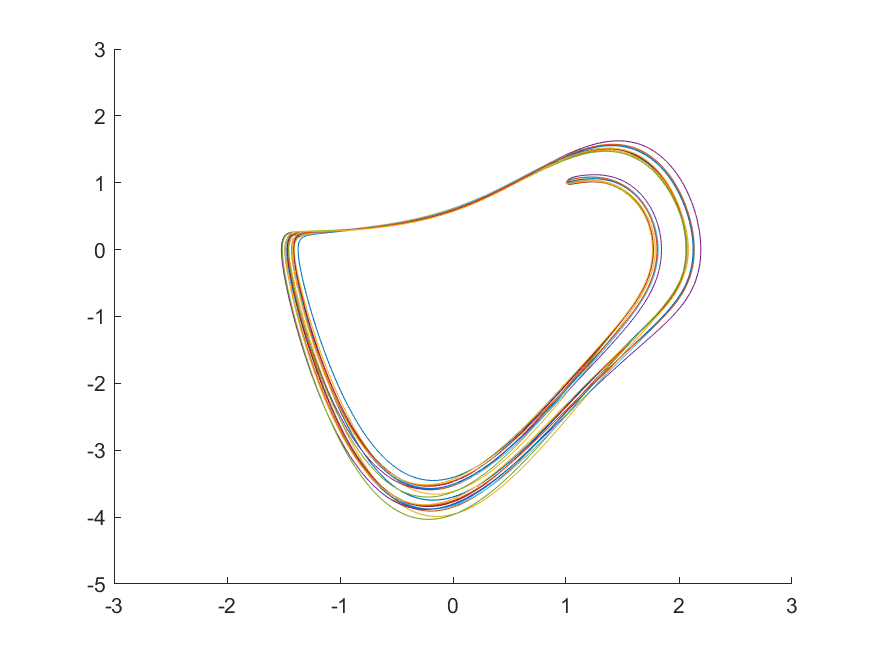}}
\newline
\subfloat[$N=100$]{\includegraphics[width=.32\textwidth]{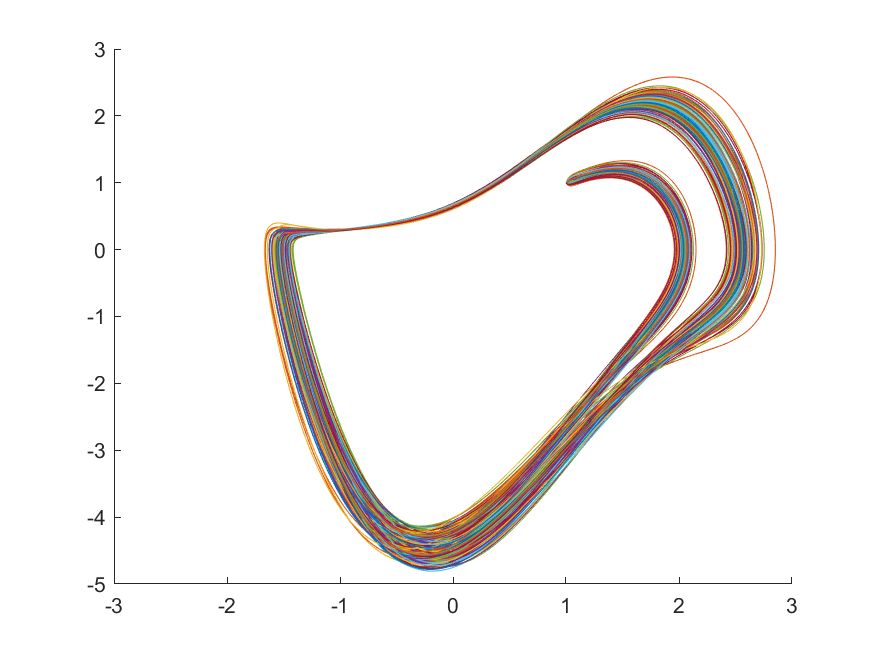}}
\subfloat[$N=100$]{\includegraphics[width=.32\textwidth]{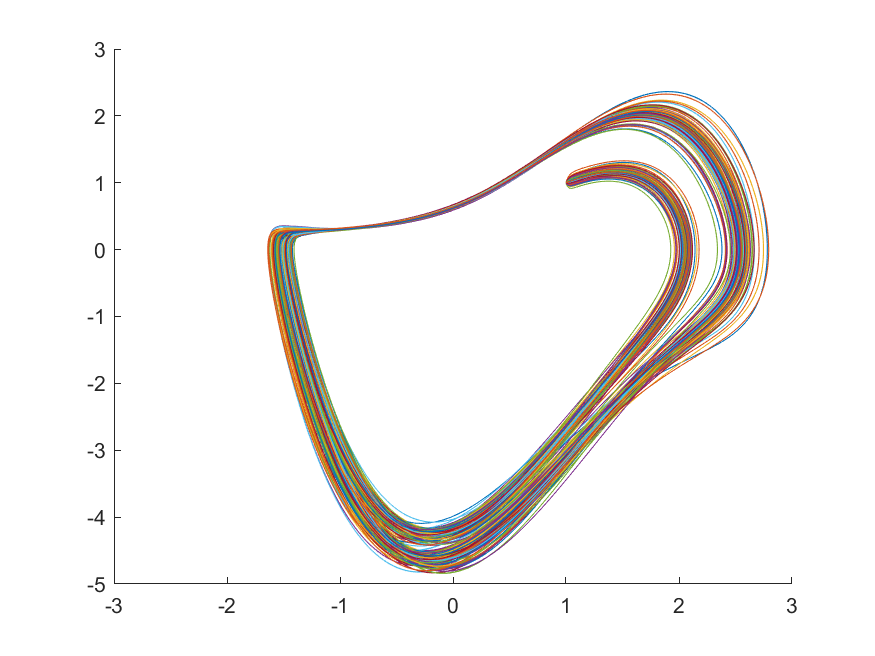}}
\subfloat[$N=100$]{\includegraphics[width=.32\textwidth]{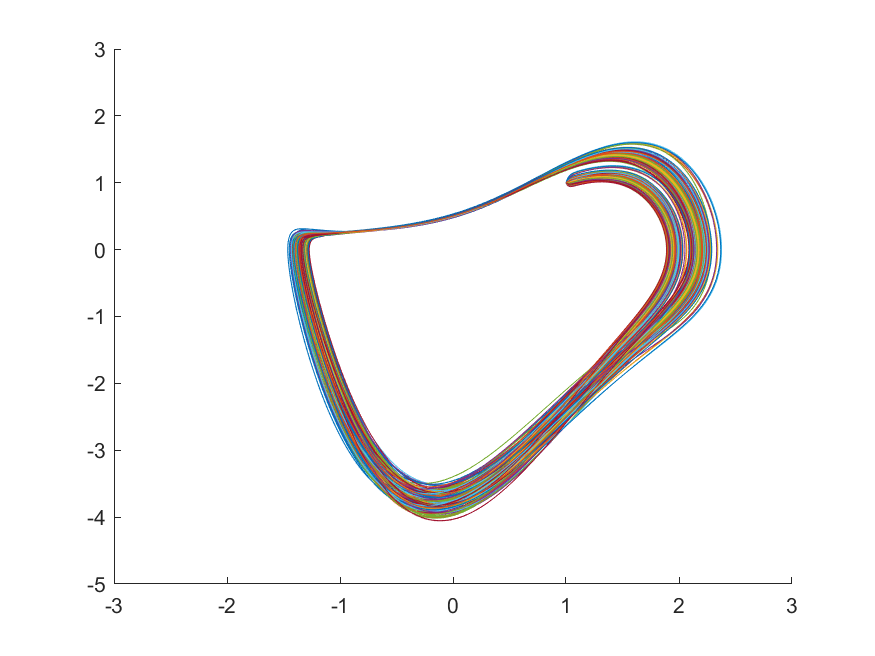}}
\newline
\subfloat[$N=1000$]{\includegraphics[width=.32\textwidth]{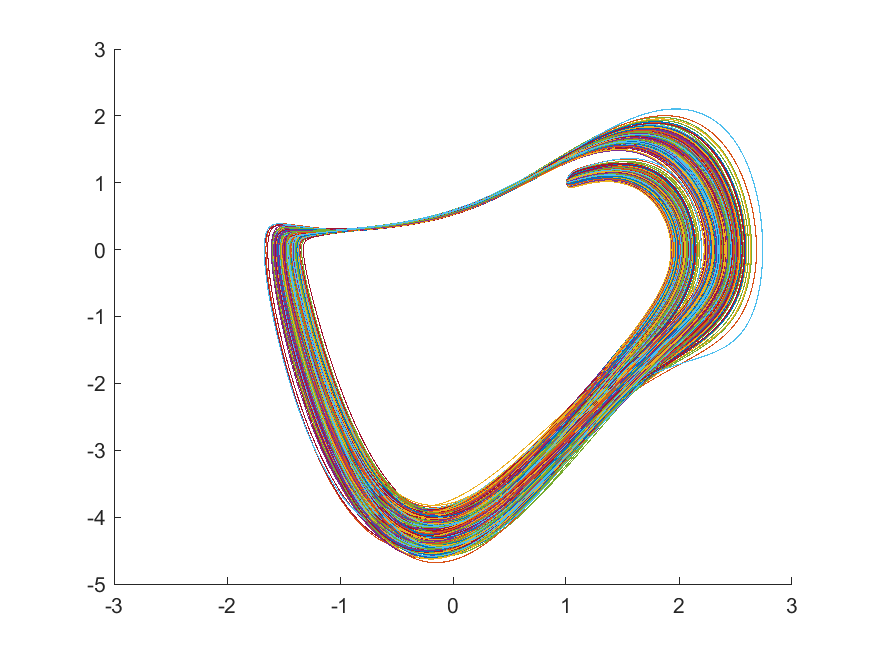}}
\subfloat[$N=1000$]{\includegraphics[width=.32\textwidth]{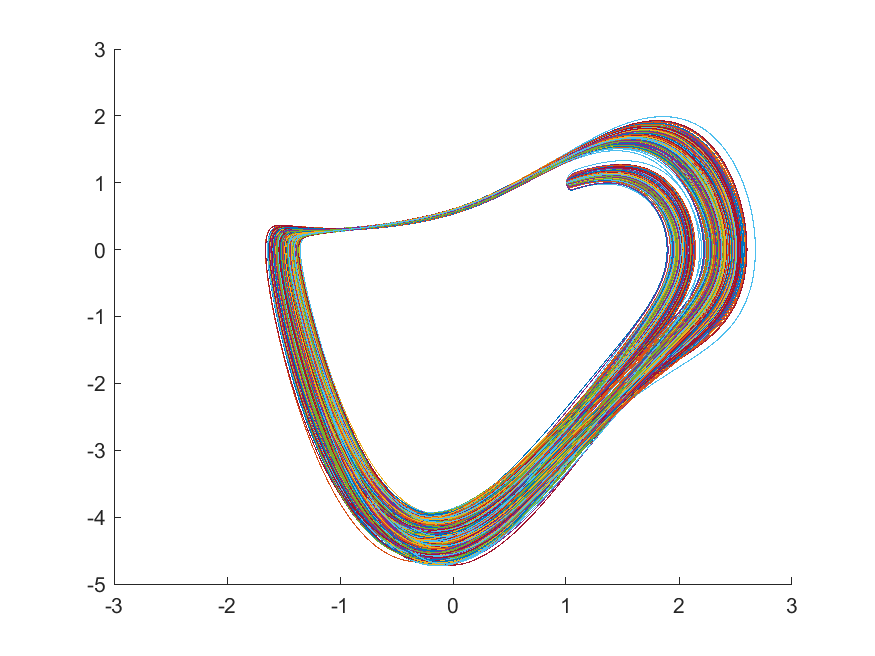}}
\subfloat[$N=1000$]{\includegraphics[width=.32\textwidth]{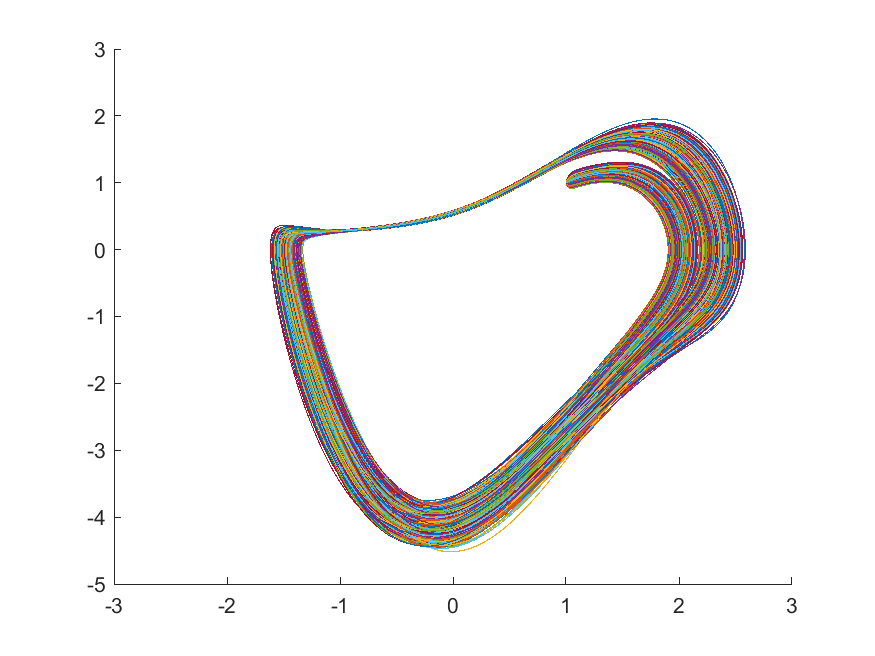}}
\caption{Simulation results of randomly generated pacemaker networks. Initial condition is
 $(1,1)$ for all cases.}
\label{fig:1}
\end{figure}

Fig.~\ref{fig:1} shows the simulation results of the pacemaker network when the number of agents are 10, 100, and 1000, respectively.
For example, we randomly generated the network for $N=10$ three times independently, and plotted their behavior in Fig.~\ref{fig:1}(a), (b), and (c), respectively.
It is seen that the variation is large in this case and Fig.~\ref{fig:1}(b) even shows the case that no stable limit cycle exists.
On the other hand, in the case when $N=1000$, the randomly generated networks exhibit rather uniform behavior as in Fig.~\ref{fig:1}(g), (h), and (i).
For the simulation, the initial condition is $z_i(0)=\dot z_i(0)=1$, the coupling gain is $k=50$, and the graph has all-to-all connection.

We refer the reader to \cite{kim2016robustness} for more discussions in this direction.

\section{More than linear coupling}

Until now, we have considered linear diffusive couplings with constant strength $k$.
In this section, let us consider two particular nonlinear couplings; edge-wise and node-wise funnel couplings, whose coupling strength varies as a nonlinear function of time and the differences between the states.

\subsection{Edge-wise funnel coupling}\label{subsec:edge}

The coupling law to be considered is inspired by the so-called funnel controller \cite{ilchmann2002tracking}.
For the multi-agent system
$$\dot x_i = f_i(t,x_i) + u_i \quad \in \R, \quad i \in \NN,$$
let us consider the following \emph{edge-wise funnel coupling} law, with $\nu_{ij} := x_j - x_i$,
\begin{align}\label{eq:funnel_coup}
u_i(t, \{\nu_{ij}, j \in \NN_i\}) &:= \sum_{j \in \mathcal{N}_i} \gamma_{ij}\left(\frac{|\nu_{ij}|}{\psi_{ij}(t)}\right)\frac{\nu_{ij}}{\psi_{ij}(t)}
\end{align}
where each function $\psi_{ij}: [t_0, \infty) \to \mathbb{R}_{> 0}$ is positive, bounded, and differentiable with bounded derivative, and the gain functions $\gamma_{ij} : [0, 1) \to \mathbb{R}_{\ge 0}$ are strictly increasing and unbounded as $s \to 1$.
We assume the symmetry of coupling functions; that is, $\psi_{ij} = \psi_{ji}$ and $\gamma_{ij}=\gamma_{ji}$ for all $i \in \mathcal{N}$ and $j \in \mathcal{N}_i$ (or, equivalently, $j \in \NN$, $i \in \NN_j$ because of the symmetry of the graph).
A possible choice for $\gamma_{ij}$ and $\psi_{ij}$ is
$$\gamma_{ij}(s) = \frac{1}{1-s} \quad \text{ and } \quad \psi_{ij}(t) = (\overline{\psi}-\eta) e^{-\lambda (t-t_0)} + \eta,$$
where $\overline{\psi},\lambda, \eta > 0$.

\begin{figure}[b]
\centering
\begin{tikzpicture}[baseline = 0,xscale=0.7,yscale=0.4,variable=\t,domain=0:10,samples=50]
   \fill[fill = blue!10] plot ({10-\t},{2/exp(0.6*(10-\t))+0.5}) -- plot (\t,{-2/exp(0.6*\t)-0.5});
   \draw[draw=black,->]  (0,-3) -- (0,3);
   \draw[draw=black,->]  (-0.5,0) -- (10.2,0) node[anchor = west] {$t$};
   \draw (12,0) node {};
   \draw[color=blue,very thick] plot (\t,{2/exp(0.6*\t)+0.5}) node[anchor = south east] {$\psi_{ij}(t)$};
   \draw[color=blue,very thick] plot (\t,{-2/exp(0.6*\t)-0.5}) node[anchor = north east] {$-\psi_{ij}(t)$};
   \draw[errorstyle] plot[smooth] coordinates {(0,1) (0.5,1.1) (1,0.9) (1.5,0.8) (2,0.7) (2.5,0.7) (3,0.75) (3.5,0.6) (4,0.4) (4.5,0.45) (5,0.1) (5.5,-0.2) (6,-0.35) (6.5,-0.4) (7,-0.45) (7.5,-0.3) (8,-0.3) (8.5,-0.2) (9,0) (9.5,0.2) (10,0.25)};
   \draw[blue] (1,-0.7) node {$\mathcal{F}_{\psi_{ij}}$};
   \draw (0,1)  node[red,left] {$\nu_{ij}(t)$};
   \end{tikzpicture}
\caption{State difference $\nu_{ij}$ evolves within the funnel $\mathcal{F}_{\psi_{ij}}$ so that the synchronization error can be prescribed by the shape of the funnel.}\label{fig:funnel}
\end{figure}
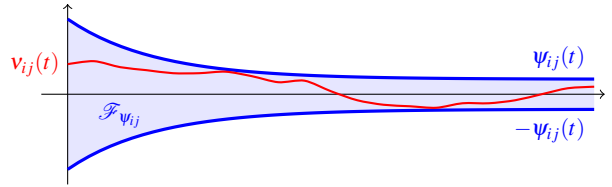

With the funnel coupling \eqref{eq:funnel_coup}, it is shown in \cite{jgleeECC} that, under the assumption that no finite escape time exists, the state difference $\nu_{ij}(t)$ evolves within the funnel:
$$\mathcal{F}_{\psi_{ij}} := \left\{(t,\nu_{ij}): |\nu_{ij}|<\psi_{ij}(t)\right\}$$
if $|\nu_{ij}(t_0)| < \psi_{ij}(t_0)$, $\forall i \in \NN$, $j \in \NN_i$, as can be seen in Fig.~\ref{fig:funnel}.
Therefore, approximate synchronization of arbitrary precision can be achieved with arbitrarily small $\eta>0$ such that $\limsup_{t\to\infty}\psi_{ij}(t) \leq \eta$.
Indeed, due to the connectivity of the graph, it follows from $\limsup_{t \to \infty} |\nu_{ij}(t)| \le \eta$, $\forall i \in \NN$, $j \in \NN_i$, that
\begin{equation}\label{eq:appsync}
\limsup_{t \to \infty} |x_j(t) - x_i(t)| \le d\eta, \qquad \forall i, j \in \NN
\end{equation}
where $d$ is the diameter of the graph.
For the complete graph, we have $d = 1$.

Here, we note that, by the symmetry of $\psi_{ij}$ and $\gamma_{ij}$ and by the symmetry of the graph, it holds that
\begin{align*}
\sum_{i=1}^N u_i &= \sum_{i=1}^N \sum_{j \in \NN_i} \gamma_{ij}\left(\frac{|\nu_{ij}|}{\psi_{ij}(t)}\right) \frac{\nu_{ij}}{\psi_{ij}} = - \sum_{i=1}^N \sum_{j \in \NN_i} \gamma_{ji}\left(\frac{|\nu_{ji}|}{\psi_{ji}(t)}\right) \frac{\nu_{ji}}{\psi_{ji}} \\
&= - \sum_{j=1}^N \sum_{i \in \NN_j} \gamma_{ji}\left(\frac{|\nu_{ji}|}{\psi_{ji}(t)}\right) \frac{\nu_{ji}}{\psi_{ji}} = - \sum_{j=1}^N u_j.
\end{align*}
Therefore, we have that
\begin{equation}\label{eq:sss}
0 = \sum_{i=1}^N u_i = \sum_{i=1}^N (\dot x_i(t) - f_i(t,x_i(t)))
\end{equation}
which holds regardless whether synchronization is achieved or not.
If all $x_i$'s are synchronized whatsoever such that $x_i(t) = s(t)$ by a common trajectory $s(t)$, it implies that $\dot x_i = f_i(t,s) + u_i = \dot s$ for all $i \in \NN$; i.e., $u_i(t)$ compensates the term $f_i(t,s(t))$ so that all $\dot x_i(t)$'s become the same $\dot s$.
Hence, \eqref{eq:sss} implies that 
\begin{equation}\label{eq:newbl}
\dot s = \frac1N \sum_{i=1}^N f_i(t,s) =: f_s(t,s).
\end{equation}
In other words, enforcing synchronization under the condition \eqref{eq:sss} yields an emergent behavior for $x_i(t) = s(t)$, governed by the blended dynamics \eqref{eq:newbl}.
In practice, the funnel coupling \eqref{eq:funnel_coup} enforces approximate synchronization as in \eqref{eq:appsync}, and thus, the behavior of the network is not exactly the same as \eqref{eq:newbl} but can be shown to be close to it.
More details are found in \cite{jgleeECC}.

A utility of the edge-wise funnel coupling is for the decentralized design.
It is because the common gain $k$, whose threshold $k^*$ contains all the information about the graph and the individual vector fields of the agents, is not used.
Therefore, the individual agent can self-construct their own dynamics when joining the network.
(For example, if it is used for the distributed least-squares solver in Section \ref{subsec:lss}, then the agent dynamics \eqref{eq:dls} can be constructed without any global information.)
Indeed, when an agent joins the network, the agent can handshake with the agents to be connected, and communicate to set the function $\psi_{ij}(t)$ so that the state difference $\nu_{ij}$ at the moment of joining resides inside the funnel.

\subsection{Node-wise funnel coupling}

Motivated by the observation in the previous subsection that the enforced synchronization under the condition \eqref{eq:sss} gives rise to the emergent behavior of \eqref{eq:newbl}, let us illustrate how different nonlinear couplings may yield different emergent behavior.
As a particular example, we consider the {\em node-wise funnel coupling} given by
\begin{equation}\label{eq:funnel_coup2}
u_i(t, \nu_i) := \gamma_i\left(\frac{|\nu_i|}{\psi_i(t)}\right)\frac{\nu_i}{\psi_i(t)} \quad \text{where} \quad \nu_i = \sum_{j \in \mathcal{N}_i} \alpha_{ij}(x_j - x_i)
\end{equation}
where each function $\psi_{i}: [t_0, \infty) \to \mathbb{R}_{> 0}$ is positive, bounded, and differentiable with bounded derivative, and the gain functions $\gamma_{i} : [0, 1) \to \mathbb{R}_{\ge 0}$ are strictly increasing and unbounded as $s \to 1$.
A possible choice for $\gamma_i$ and $\psi_i$ is
\begin{equation}\label{eq:sample}
\gamma_i(s) = \begin{cases} \frac{\delta}{s} \tan \left(\frac{\pi}{2}s\right), & s > 0 \\ \frac{\pi}{2}\delta, & s = 0 \end{cases} \qquad \text{and} \quad \psi_i(t) = (\overline{\psi} - \eta)e^{-\lambda (t-t_0)} + \eta
\end{equation}
where $\delta, \overline{\psi}, \lambda, \eta > 0$.

With the funnel coupling \eqref{eq:funnel_coup2}, it is shown in \cite{lee2019synchronization} that, under the assumption that no finite escape time exists, the quantity $\nu_{i}(t)$ evolves within the funnel $\mathcal{F}_{\psi_{i}} := \left\{(t,\nu_{i}): |\nu_{i}|<\psi_{i}(t)\right\}$ if $|\nu_{i}(t_0)| < \psi_{i}(t_0)$, $\forall i \in \NN$.
Therefore, approximate synchronization of arbitrary precision can be achieved with arbitrarily small $\eta>0$ such that $\limsup_{t\to\infty}\psi_{i}(t) \leq \eta$.
Indeed, due to the connectivity of the graph, it follows that
\begin{equation}\label{eq:appsync2}
\limsup_{t \to \infty} |x_j(t) - x_i(t)| \le \frac{2\sqrt{N}}{\lambda_2}\eta, \qquad \forall i, j \in \NN 
\end{equation}
where $\lambda_2$ is the second smallest eigenvalue of $\LL$.

Unlike the case of edge-wise funnel coupling, there is lack of symmetry so that the equality of \eqref{eq:sss} does not hold.
However, assuming that the map $u_i(t,\nu_i)$ from $\nu_i$ to $u_i$ is invertible (which is the case for \eqref{eq:sample} for example) such that there is a function $V_i$ such that 
$$\nu_i = V_i(t,u_i(t,\nu_i)), \qquad \forall t, \nu_i,$$
we can instead make use of the symmetry in $\nu_i$ as
$$\sum_{i=1}^N \sum_{j \in \NN_i} \alpha_{ij} (x_j-x_i) = - \sum_{i=1}^N \sum_{j \in \NN_i} \alpha_{ji} (x_i-x_j) = - \sum_{j=1}^N \sum_{i \in \NN_j} \alpha_{ji} (x_i-x_j),$$
which leads to
\begin{equation}\label{eq:sss2}
0 = \sum_{i=1}^N \nu_i = \sum_{i=1}^N V_i(t,u_i(t,\nu_i)) = \sum_{i=1}^N V_i(t,\dot x_i(t) - f_i(t,x_i(t))).
\end{equation}
This holds regardless whether synchronization is achieved or not.
If all $x_i$'s are synchronized whatsoever such that $x_i(t) = s(t)$ by a common trajectory $s(t)$, it implies that $\dot x_i(t) = f_i(t,s) + u_i(t) = \dot s$ for all $i \in \NN$; i.e., $u_i(t)$ compensates the term $f_i(t,s)$ so that all $\dot x_i$ are the same as $\dot s$, which can be denoted by $f_s(t,s) = f_i(t,s)+u_i(t)$.
Hence, \eqref{eq:sss2} implies that 
\begin{equation}\label{eq:newbl2}
\sum_{i=1}^N V_i(t,f_s(t,s) - f_i(t,s)) = 0.
\end{equation}
In other words, \eqref{eq:newbl2} defines $f_s(t,s)$ implicitly, which yields the emergent behavior governed by
\begin{equation}\label{eq:newbl3}
\dot s = f_s(t,s).
\end{equation}
In practice, the funnel coupling \eqref{eq:funnel_coup2} enforces approximate synchronization as in \eqref{eq:appsync2}, and the behavior of the network is not exactly the same as \eqref{eq:newbl3} but it is shown in \cite{lee2019synchronization} to be close to \eqref{eq:newbl3}.

In order to illustrate that different emergent behavior may arise from various nonlinear couplings, let us consider the example of \eqref{eq:sample}, for which the function $V_i$ is given by
$$V_i(t,u_i) = \frac{2\psi_i(t)}{\pi} \tan^{-1}\left(\frac{u_i}{\delta}\right).$$
Assuming that all $\psi_i$'s are the same, the emergent behavior $\dot{s} = f_s(t, s)$ can be found with $f_s(t, s)$ being the solution to
$$0 = \sum_{i=1}^N \tan^{-1}\left(\frac{f_s(t, s) - f_i(t, s)}{\delta}\right).$$
If we let $\delta \to 0$, then the above equality shares the solution with
$$0 = \sum_{i=1}^N \text{sgn}\left(f_s(t, s) - f_i(t, s)\right).$$
Recalling the discussions in Section \ref{subsec:med}, it can be shown that $f_s(t, s)$ takes the median of all the individual vector fields $f_i(t, s)$, $i=1,\dots, N$.
Since taking median is a simple and effective way to reject outliers, this observation may find further applications in practice.

\section*{Acknowledgement}

This work was supported by the National Research Foundation of Korea grant funded by the Korea government (Ministry of Science and ICT) under No.~NRF-2017R1E1A1A03070342 and No.~2019R1A6A3A12032482.
This is a preprint of the following chapter: Jin Gyu Lee and Hyungbo Shim, ``Design of heterogeneous multi-agent system for distributed computation,'' published in Trends in Nonlinear and Adaptive Control, edited by Zhong-Ping Jiang, Christophe Prieur, and Alessandro Astolfi, 2021, Springer reproduced with permission of Springer.
The final authenticated version is available online at: http://dx.doi.org/10.1007/978-3-030-74628-5\_4

\end{document}